\documentclass[reprint,amsmath,amssymb,floatfix]{revtex4-2}
\usepackage{geometry}

\usepackage{amsmath}
\usepackage{amsfonts}
\usepackage{amssymb}
\usepackage{graphicx}
\usepackage{dcolumn}
\usepackage{bm}
\usepackage[colorlinks = true, allcolors = blue]{hyperref}
\usepackage{float}

\makeatletter
\def\ps@pprintTitle{%
	\let\@oddhead\@empty
	\let\@evenhead\@empty
	\def\@oddfoot{}%
	\let\@evenfoot\@oddfoot}
\makeatother
\begin{document}
	\definecolor{background-color}{gray}{0.98}
	
	\newcommand{\bea}{\begin{eqnarray}}
		\newcommand{\eea}{\end{eqnarray}}
	\newcommand{\bes}{\begin{subequations}}
		\newcommand{\ees}{\end{subequations}}
	\newtheorem{dfn}{Definition}[section]
	\newtheorem{ex}{Example}[section]
	\newtheorem{subex}{Example}[subsection]
	\newtheorem{cl}{Corrolary}[section]
	\newtheorem{propo}{Proposition}[section]
	\newtheorem{theorem}{Theorem}[section]

	\newcommand{\bd}{\begin{document}}
		\newcommand{\ed}{\end{document}}
	\newcommand{\bc}{\begin{center}}
		\newcommand{\ec}{\end{center}}
	\newcommand{\bfr}{\begin{flushright}}
		\newcommand{\efr}{\end{flushright}}
	\newcommand{\lt}{\left}
	\newcommand{\rt}{\right}
	\newcommand{\vs}{\vspace}
	\newcommand{\hs}{\hspace}
	\newcommand{\beq}{\begin{equation}}
		\newcommand{\eeq}{\end{equation}}
	\newcommand{\lb}{\linebreak}
	\newcommand{\pb}{\pagebreak}
	\newcommand{\mb}{\makebox}
	\newcommand{\fb}{\framebox}
	\newcommand{\mc}{\multicolumn}
	\newcommand{\ben}{\begin{enumerate}}
		\newcommand{\een}{\end{enumerate}}
	\newcommand{\bit}{\begin{itemize}}
		\newcommand{\eit}{\end{itemize}}
	\newcommand{\oln}{\overline}
	\newcommand{\un}{\underline}
	\newcommand{\lefq}{\lefteqn}
	\newcommand{\ba}{\begin{array}}
		\newcommand{\ea}{\end{array}}
	\newcommand{\beqa}{\begin{eqnarray}}
		\newcommand{\eeqa}{\end{eqnarray}}
	\newcommand{\beqas}{\begin{eqnarray*}}
		\newcommand{\eeqas}{\end{eqnarray*}}
	\newcommand{\bfg}{\begin{figure}}
		\newcommand{\efg}{\end{figure}}
	\newcommand{\bds}{\begin{displaymath}}
		\newcommand{\eds}{\end{displaymath}}
	\newcommand{\btb}{\begin{tabbing}}
		\newcommand{\etb}{\end{tabbing}}
	\newcommand{\para}{\parallel}
	\newcommand{\pad}{\partial}
	\newcommand{\nn}{\nonumber}
	\newcommand{\la}{\leftarrow}
	\newcommand{\ra}{\rightarrow}
	\newcommand{\lgla}{\longleftarrow}
	\newcommand{\lgra}{\longrightarrow}
	\newcommand{\La}{\Leftarrow}\newcommand{\Ra}{\Rightarrow}
	\newcommand{\Lra}{\Leftrightarrow}
	\newcommand{\Lgla}{\Longleftarrow}
	\newcommand{\Lgra}{\Longrightarrow}
	\newcommand{\lan}{\langle}
	\newcommand{\ran}{\rangle}
	\renewcommand{\a}{\alpha}
	\renewcommand{\b}{\beta}
	\newcommand{\g}{\gamma}
	\newcommand{\G}{\Gamma}
	\renewcommand{\d}{\delta}
	\newcommand{\eps}{\epsilon}
	\newcommand{\Th}{\Theta}
	\newcommand{\s}{\sigma}
	\newcommand{\lam}{\lambda}
	\newcommand{\D}{\Delta}
	\newcommand{\ds}{\displaystyle}
	\newcommand{\vare}{E}
	\newcommand{\pr}{\prime}
	\newcommand{\ro}{\rho}
	\newcommand{\nab}{\nabla}
	\newcommand{\m}{\mu}
	\newcommand{\n}{\nu}
	\newcommand{\Sg}{\Sigma}
	\newcommand{\p}{\pi}
	\newcommand{\R}{I\!\!R}
	\newcommand{\om}{\omega}
	\newcommand{\Om}{\Omega}
	\newcommand{\ovra}{\overrightarrow}
	\newcommand{\ze}{\zeta}
	\newcommand{\vart}{\vartheta}
	\newcommand{\tri}{\triangle}
	\newcommand{\f}{\frac}
	\newcommand{\iny}{\infty}
	\newcommand{\pro}{\propto}
	\renewcommand{\arraystretch}{1.25}
	\title{Majorization effect on entropic functional: An application to a $\vee$-type three-level atom-field interacting system}
	\author{Debraj Nath}
	\affiliation{Department of Mathematics, Vivekananda College, Kolkata-700063, WB, India.}
	\email{debrajn@gmail.com} 	
	
	\begin{abstract}
		Majorization effect on some entropic functional, such as von-Neumann, Shannon, atomic Wehrl and R\'enyi entropies are investigated of a $\vee$-type three-level atom, which interacts with a coherent field in a resonant cavity. Fidelity, purity and linear entropy are investigated for two quantum systems, which contain photon number distributions and Husimi $Q$ functions. A relation between majorization and localization properties of continuous density functional is established. The results are compared and verified for continuous and discrete distributions of an atom-field interacting system.
	\end{abstract}
	\maketitle
	\section{Introduction}
	The interaction between a two-level atom and a single-mode quantized electromagnetic field in a lossless cavity is known as Jaynes-Cummings model (JCM) \cite{jcm}. It is exactly solvable under the rotating-wave approximation. It has many applications in quantum optics, such as Rabi oscillations, atomic inversion, sub-Poissonian statistics, squeezing and phase properties of radiation field. Moreover, the interaction between three-level atom(s) and field(s) have an important role in quantum optics, in the context of collapse-revival phenomena, squeezing phenomena, entanglement, quantum information, quantum computation and so on. A large number of theoretical and numerical studies have been made on the subject of quantum information entropy of a three-level atom of ladder, vee and lambda configurations, which interacts with single-mode and multi-mode quantized field \cite{tla.information,tla.information.3,tla.information.4,tla.information.5.4,tla.information.6,tla.information.8,tla.information.9,TLA.vee,TLA.vee.2,TLA.vee.3,TLA.vee.4,TLA.vee.5,TLA.vee.6,TLA.vee.7,TLA.vee.8,TLA.vee.9,tla,tla.information2.4,tla.information2.5,tla.information2.7,x}. In this connection, we have noted that, different approaches have been used to determine the population of the excited states of a three-level atom \cite{tla.information,tla.information.3,tla.information.4,tla.information.5.4,tla.information.6,tla.information.8,tla.information.9,TLA.vee,TLA.vee.2,TLA.vee.3,TLA.vee.4,TLA.vee.5,TLA.vee.6,TLA.vee.7,TLA.vee.8,TLA.vee.9,tla}, which is a measure of the two-photon absorption probability \cite{absorption}. These are associated with a two-level atom, driven by a single-mode field and they are extension of Jaynes-Cummings model \cite{jcm}. If an atom collapses to the ground state with a probability $p$, then the probability $p$ is called fidelity \cite{fidelity1,Fidelity2}. In this paper, we have consider the maximum value of $p$. A quantum state with density operator $\rho$, is said to be pure, if $\rho^2=\rho$. The measure of purity $Tr[\rho^2]$ of a moving three-level atom is investigated in ref. \cite{tla.information.6}. From the definition of purity, one can define the linear entropy $1-Tr[\rho^2]$ as a complement measure of purity of a state with normalized density functions $\rho$ \cite{pipek,pipek.2}, which is obtained from the von-Neumann entropy in a near pure state by Taylor series expansion. The linear entropy of tripartite two-level atom \cite{pipek.2.2}, three-level atom \cite{tla.information.5.4} and four-level atom \cite{pipek.3} were investigated earlier in different context. Many authors focus their attention for measure the von-Neumann entropy \cite{v1} of three-level atom(s) field interacting quantum system \cite{tla.information,tla.information.3,tla.information.4,tla.information.5.4,tla.information.6,tla.information.8,tla.information.9,tla.information2.4,tla.information2.5,tla.information2.7,x}. It is an important measurement in the field of quantum information and computation for mixed states. The von-Neumann entropy of an atom and a field are investigated in a dissipative cavity in ref. \cite{tla.information.4} and in the presence of non-linearity of a three-level atom of different configurations, in ref. \cite{TLA.vee.6}. Another important entropy is Shannon entropy \cite{S1} which is used for the measure of temporal evolution of a field in an atom-field interacting system  (see refs. \cite{pln.aop,orlowski}). The Shannon entropy investigated for three-level atom system (see refs. \cite{tla.information.4,tla.information.8}).
	
	In this paper, a vee-type three-level atom have been considered, which interacts with a field in a resonant cavity \cite{TLA.vee, TLA.vee.2, TLA.vee.3, TLA.vee.4, TLA.vee.5, TLA.vee.6, TLA.vee.7, TLA.vee.8, TLA.vee.9}. We would like to study the atomic coherent state \cite{acs} of a three-level atom, whose evolution started with a coherent super position of its excited states and it is interacting with a coherent field. The Husimi distribution is the expectation of a quantum state and the expectation is defined by a basis of coherent states \cite{husimi}, which are defined on a complex plane. The Boltzmann-Gibbs entropy of the Husimi distribution is introduced by Wehrl and it is called the atomic Wehrl entropy. It is minimum for coherent state and is conjectured by Wehrl \cite{w1}. After that, it is proved by Lieb \cite{lieb}. It is a measure of coherence of a given state. The atomic Wehrl entropy of three-level atom(s) is investigated in \cite{tla.information.4}. The R\'enyi entropy is a generalization of Shannon entropy of arbitrary order \cite{Renyi}. It has many applications in the fields of statistical mechanics, quantum information and computation. 
	
	Generally, majorization properties are used to compare the probability distributions. Moreover, majorization techniques are used in quantum information specially in quantum entanglement \cite{Lewenstein}. If two distributions can compare, then their corresponding entropic uncertainty measurements can be compared. In addition, the entropic uncertainty relations for any two observables can be established by the majorization techniques \cite{majo.delgado,majo.delgado2,majorization.open,Lewenstein,majorization}. The lower limit of the R\'enyi entropic sum for bipartite Hilbert space is established in ref. \cite{Lewenstein}. In most of the published papers majorization technique has been shown for discrete distributions.
	
	In this paper, the majorization and the cumulative distribution of discrete and continuous density functions will be described. Under this model, there are two set of interacting quantum systems (i) first atom-field and (ii) second atom-field. Both systems contain photon number distributions and quasi probability distribution. The purity, linear entropy and the von-Neumann entropy of reduced density matrices of atoms will be defined. Then, the Shannon and R\'enyi entropies will be defined of photon number distributions whereas, the atomic Wehrl and the R\'enyi entropies will be defined of quasi probability distribution (Husimi $Q$-function). To understand the entropic nature, the properties of majorization and cumulative distributions will be investigated. Moreover, effective domain of continuous probability density function will be defined. Then characterize its localization property with respect to the effective domain \cite{IJQC.local,dn.rcr}. Next, the connection between majorization and localization property of continuous density functions will be investigated. The motivation of this paper is to find the majorization effect on some entropic functional. As an example, we have considered an atom-field interacting system. In this regard, the possibilities for reducing values of information theoretic measures, such as purity, linear entropy \cite{pipek,pipek.2,pipek.3}, von-Neumann entropy \cite{v1}, Shannon entropy \cite{S1}, atomic Wehrl entropy \cite{w1,w2,w2.2,w2.3} and R\'enyi entropy \cite{Renyi} will be examined by adding another identical atom into the cavity, when the first atom collapses to the ground state, after some time. The time will be taken, in such a way that, the atom collapses to the ground state with maximum fidelity \cite{fidelity1} and then the atom-field interacting system will reduce to a single-mode field and thereafter the second atom will be passed through the cavity. The detail calculations of information measures depend on the initial conditions of the atom and the field. 
	
	The article is organized as follows. In Section \ref{sec2.model}, a theoretical model of an atom-field interaction in a cavity of quantized electromagnetic field will be defined. Then, exact solutions of this model for two quantum systems, such as (i) first atom-field interaction and (ii) second atom-field interaction will be calculated. In Section \ref{sec3.preliminaries}, some definitions of entropies, such as von-Neumann, Shannon, atomic Wehrl and R\'enyi will be reviewed. In addition, a relation between majorization and localization property of density functions will be defined. In Section \ref{sec4.application}, a application of entropic functional to this model will be demonstrated. Numerical results of entropic functional will be compared and verified of a vee type three-level atom in two quantum systems. Finally, in Section \ref{sec5.conclusion}, some conclusions will be exhibited.
	
	\section{Theoretical model}\label{sec2.model}
\begin{figure}[h!]
		\centering
		\includegraphics[width=.3\textwidth]{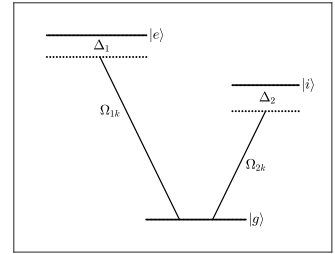}
		\caption{\label{Fig1} Schematic representation of a vee-type three-level atom. $\omega$ is carried frequency of the driving field and $\Delta_{1}$, $\Delta_{2}$ are the detuning parameters.}
	\end{figure}

	As an application, we have considered a three-level vee type atom \cite{TLA.vee,TLA.vee.2,TLA.vee.3,TLA.vee.4,TLA.vee.5,TLA.vee.6,TLA.vee.7,TLA.vee.8,TLA.vee.9} in a radiation field as shown in Fig. \ref{Fig1}. The allowed photon transitions are $\ds|e\rangle\leftrightarrow|g\rangle$ and $\ds|i\rangle\leftrightarrow|g\rangle$, where $|e\rangle$, $|i\rangle$, $|g\rangle$ are eigenstates of a three-level atom with eigenenergies $\omega_{e}$, $\omega_{i}$, $\omega_{g}$. For, $\vee$-type three level atom, there is no direct photon transition between levels $|i\rangle$ and $|e\rangle$. But it can be allowed for $\Lambda$ and $\varXi$ and $\Delta$ types three-level atom. For the sake of simplicity, a linear polarization basis is taken and the corresponding polarization unit vector is taken as real. Then the total Hamiltonian of an interacting system of a quantized electromagnetic field and a vee type three-level atom is,
	\begin{equation}\label{Total.Hamiltonian}
		\begin{array}{ll}
			H&=\sum\limits_{k}\hbar\left(\Om_{1k}a_{k}^{\dag}a_{k}+\Om_{2k}b_{k}^{\dag}b_{k}\right)+\sum\limits_{\mu=e,i,g}\omega_{\mu}|\mu\rangle\langle\mu|\\
			&+\hbar\sum\limits_{k}\left(\sum\limits_{l,m=e,g}g_{1k}^{lm}|l\rangle\langle m|\left(a_{k}+a^{\dag}_{k}\right)\right.\\
			&\left.+\sum\limits_{l,m=i,g}g_{2k}^{lm}|l\rangle\langle m|\left(b_{k}+b^{\dag}_{k}\right)\right)\\
		\end{array}
	\end{equation} 
	where $a^{\dag}$ and $b^{\dag}$ (or $a$ and $b$) are creation (or annihilation) operators of the quantized driving fields with frequencies $\Om_{1k}$ and $\Om_{2k}$ respectively. The terms $g_{1k}^{lm}$, and $g_{2k}^{lm}$ are the coupling constants. For the resonant cavity $|e\rangle\langle g|a_{k}$, $|g\rangle\langle e|a_{k}^{\dag}$, $|i\rangle\langle g|b_{k}$ and $|g\rangle\langle i|b_{k}^{\dag}$ are energy conserving operators. The operators $|e\rangle\langle g|a_{k}^{\dag}$ and $|i\rangle\langle g|b_{k}^{\dag}$ gain the energies $\omega_e-\omega_g+\hbar\Om_{1k}$ and $\omega_i-\omega_g+\hbar\Om_{2k}$ respectively. Similarly, the operators $|g\rangle\langle e|a_{k}$ and $|g\rangle\langle i|b_{k}$ loss the energies $\omega_e-\omega_g+\hbar\Om_{1k}$ and $\omega_i-\omega_g+\hbar\Om_{2k}$ respectively. Therefore, the non conserving operators have been neglected. In the rotating-wave approximation, the total Hamiltonian is written by
	\begin{equation}
		\begin{array}{ll}
			H&=\sum_{k}\hbar\left(\Om_{1k}a_{k}^{\dag}a_{k}+\Om_{2k}b_{k}^{\dag}b_{k}\right)+\sum_{\mu=e,i,g}\omega_{\mu}|\mu\rangle\langle\mu|\\
			&+\hbar\sum_{k}g_{1k}^{eg}\left(|e\rangle\langle g|a_{k}+|g\rangle\langle e|a^{\dag}_{k}\right)\\
			&+\hbar\sum_{k}g_{2k}^{ig}\left(|i\rangle\langle g|b_{k}+|g\rangle\langle i|b^{\dag}_{k}\right).
		\end{array}
	\end{equation} 
	Therefore, for a single-mode quantized field the total Hamiltonian reduces to 
	\beq\label{Hamiltonian}
	\ds H=H_{0}+H_{1},
	\eeq
	where
	\beq
	\ba{rcl}
	\ds H_{0}& = &\omega a^{\dag}a+\omega_{g}|g\rangle \langle g| +\omega_{i}|i\rangle \langle i| + \omega_{e}|e\rangle \langle e|,\\
	\ds H_{1}& = &g_{1}\left(~a|e\rangle \langle g| + a^{\dag}|g\rangle \langle e|~\right) + g_{2}\left(~a|i\rangle \langle g| + a^{\dag}|g\rangle \langle i|~\right),
	\ea
	\eeq
	$a^{\dag}$ (or $a$) is the creation (or annihilation) operator of the quantized field with frequency $\omega$ and $g_{1}$, $g_{2}$ are the corresponding atom-field coupling constants. For the sake of simplicity, real coupling constant is considered and $\hbar=1$. Then the interaction Hamiltonian becomes
	\beq\label{S.pic}
	\ba{rcl}
	\mathcal{H} 
	& = & \ds g_{1}e^{i\Delta_{1}t}a|e\rangle \langle g| + g_{2}e^{i\Delta_{2}t}a|i\rangle \langle g| + H.C,
	\ea
	\eeq
	where $\ds \Delta_{1} = \omega_{e} - \omega_{g} - \omega$ and $\ds \Delta_{2} = \omega_{i} - \omega_{g} - \omega$. 
	\subsection{Analytical solution of first atom-field interacting system}
	In the interacting picture, the state vector of an atom-field system at time $t$ can be described by
	\beq\label{at1}
	\ds |\psi_{1}(t)\rangle  =\sum_{n=0}^{\infty}C^{1}_{e,n}(t)~|e, n\rangle + C^{1}_{i,n}(t)~|i, n\rangle + C^{1}_{g,n}(t)~|g, n\rangle.
	\eeq
	Using (\ref{at1}) into the Schr\"odinger equation  
	\beq\label{Schro.pic}
	\ds i\frac{\partial |\psi_{1}(t)\rangle }{\partial t}= \mathcal{H} |\psi_{1}(t)\rangle,
	\eeq
	one obtains
	\beq\label{cgie}
	\ba{l}
	\ds i \dot{C}^{1}_{g,n+1} = g_{1}e^{-i\Delta_{1}t}\sqrt{n + 1}~C^{1}_{e,n} + g_{2}e^{-i\Delta_{2}t}\sqrt{n + 1}~C^{1}_{i,n},\\
	\ds i \dot{C}^{1}_{i,n}  = g_{2}e^{i\Delta_{2}t}\sqrt{n + 1}~C^{1}_{g,n+1},\\
	\ds i \dot{C}^{1}_{e,n}  = g_{1}e^{i\Delta_{2}t}\sqrt{n + 1}~C^{1}_{g,n+1}.
	\ea
	\eeq
	In this paper, $\ds\Delta_{1} = \Delta_{2} = \Delta$ and the atom at initial time is a coherent superposition $\left(\ds |\psi_{1A}(0)\rangle = \cos\frac{\varphi}{2} |e\rangle\right.$ $\left. + \sin\frac{\varphi}{2} e^{i\psi}|i\rangle\right)$ of two excited states $|e\rangle$ and $|i\rangle$. Moreover, the field at initial time is the superposition of the photon number states $\left(\ds |\psi_{1F} (0)\rangle =\sum_{n=0}^{\infty}C_{n}|n\rangle, ~\sum_{n=0}^{\infty}|C_{n}|^{2} = 1\right)$. Then, the initial state vector of the atom field interacting system can be written as,
	\beq\label{con1}
	\ds |\psi_{1}(0)\rangle=\ds \sum_{n=0}^{\infty}C_{n}\left(\cos\frac{\varphi}{2}~|e,n\rangle + \sin\frac{\varphi}{2}e^{i\psi}~|i,n\rangle\right).
	\eeq
	Using the initial condition (\ref{con1}), one obtains the solution of (\ref{cgie}) as
	\beq
	\ba{ll}\label{first}
	\ds C^{1}_{e,n}(t) & =\frac{A_{1}g_{1}}{(g_{1}^{2}+g_{2}^{2})\sqrt{n + 1}}\left\{\mu + \left(\frac{i}{2}\Delta\sin\mu t-\mu\cos\mu t\right)e^{\frac{i}{2}\Delta t}\right\}\\
	& + \sin\frac{\varphi}{2} e^{i\psi}~C_{n},\\
	\ds C^{1}_{i,n}(t) &= \frac{A_{1}g_{2}}{(g_{1}^{2}+g_{2}^{2})\sqrt{n + 1}}\left\{\mu + \left(\frac{i}{2}\Delta\sin\mu t-\mu\cos\mu t\right)e^{\frac{i}{2}\Delta t}
	\right\}\\
	&+ \cos\frac{\varphi}{2}~C_{n},\\
	\ds C^{1}_{g,n+1}(t) &= A_{1}(n)ie^{-\frac{i}{2}\Delta t} \sin\mu t,
	\ea
	\eeq
	where
	\beq
	\ba{l}\label{amu1}
	\ds\mu =\frac{1}{2}\sqrt{\Delta^{2} + 4(g_{1}^{2} + g_{2}^{2})(n + 1)},\\
	\ea
	\eeq
	and 
	$A_1$ is a function of $n$ and it is written as 
	\beq
	A_1(n)=-\frac{1}{\mu}\left(g_{1}\sin\frac{\varphi}{2}e^{i\psi}+g_{2}\cos\frac{\varphi}{2}\right)\sqrt{n + 1}~C_{n}
	\eeq
	As an application, in section 4.1 we will considered $C_{n} = e^{-\frac{|\alpha|^{2}}{2}}\frac{\alpha^{n}}{\sqrt{n!}}$, where $\alpha=\sqrt{\bar{n}}~e^{i\zeta}$. From Eqs. (\ref{first}) and (\ref{amu1}), we obtain $C^1_{g,0}=0$. Therefore, $|g,0\ran$ will not be present in Eq. (\ref{at1}) for any other calculations. The coefficients $C^{1}_{e,n}(t)$, $C^{1}_{i,n}(t)$ and $C^{1}_{g,n}(t)$ in solution (\ref{at1}) satify the condition $\sum\limits_{n=0}^{\infty}\left[|C^{1}_{e,n}(t)|^2+|C^{1}_{i,n}(t)|^2+|C^{1}_{g,n}(t)|^2\right]=1$ and $C^{1}_{e,n}(t),C^{1}_{i,n}(t),C^{1}_{g,n}(t)\rightarrow 0$ as $n\rightarrow\infty$. In this paper, we have checked that $|C^{1}_{e,n}(t)|,|C^{1}_{i,n}(t)|,|C^{1}_{g,n}(t)|<10^{-9}$ for $n>30$ for the driving field with mean photon number is $\bar{n}=8$. In this paper, we have consider the driving field is a coherent laser field.
	
	\subsection{Fidelity }
	Let us assume that, the first atom collapses to the ground state at time $t$ with fidelity $f_1(t)$, which is defined by \cite{fidelity1} 
	\beq\label{fidelity}
	f_1(t)=\f{\ds\sum\limits_{n=0}^{\infty}|C^{1}_{g,n+1}(t)|^{2}}{\ds\sum\limits_{n=0}^{\infty}\left(|C^{1}_{e,n}(t)|^{2}+|C^{1}_{i,n}(t)|^{2}+|C^{1}_{g,n+1}(t)|^{2}\right)}.
	\eeq 
	It satisfies the relation $0\le f_1(t)\le 1$, for $t\ge 0$. The normalized time is denoted by \(\tau\) and it is defined by $\tau=gt$. Let $f_1$ be maximum at $\ds\tau_1=gt_1$ and the atom-field interacting system becomes
	\beq
	\ds|\psi_{1}(t_{1})\rangle = \eta_1\sum_{n=0}^{\infty}C^{1}_{g,n+1}(t_{1})~|n + 1\rangle,
	\eeq
	where $\ds\eta_1(t_1)=\left(\sum_{n=0}^{\infty}|C^{1}_{g,n+1}(t_{1})|^{2}\right)^{-\frac{1}{2}}$, is normalization constant and $\ds C^{1}_{g,n+1}(t_{1})$ is obtained from Eq. (\ref{first}) by setting $t = t_{1}$. 
	\subsection{Analytical solution of second atom-field interacting system}
	Now, another identical atom which passes through the cavity, when the first atom leaves the cavity \cite{gsa,gsa.2}. Then, the entangled state of the second atom-field interacting system, at time $t$, is written by,
	\beq\label{at2}
	\ds|\psi_{2}(t)\rangle  =\sum_{n=0}^{\infty}C^{2}_{e,n}(t)~|e,n\rangle + C^{2}_{i,n}(t)~|i,n\rangle + C^{2}_{g,n}(t)~|g,n\rangle.
	\eeq
	Now, substituting (\ref{at2}) into the Schr\"odinger equation (\ref{Schro.pic}), we obtain
	\beq
	\ba{l}\label{c.gie2}
	\ds i \dot{C}^{2}_{g,n+1} = g_{1}e^{-i\Delta_{1}t}\sqrt{n + 1}~C^{2}_{e,n} + g_{2}e^{-i\Delta_{2}t}\sqrt{n + 1}~C^{2}_{i,n},\\\\
	\ds i \dot{C}^{2}_{i,n} = g_{2}e^{i\Delta_{2}t}\sqrt{n + 1}~C^{2}_{g,n+1},\\\\
	i \dot{C}^{2}_{e,n} = g_{1}e^{i\Delta_{2}t}\sqrt{n + 1}~C^{2}_{g,n+1}.
	\ea
	\eeq
	In a similar manner, $\ds|\psi_{2A}(0)\rangle = \cos\frac{\varphi^{'}}{2} |e\rangle + \sin\frac{\varphi^{'}}{2}e^{i\psi^{'}}|i\rangle$. Then the state vector for the second atom-field interacting system at $t = 0$, can be written as,
	\beq\label{cond2}
	\ds|\psi_{2}(0)\rangle =\eta_1\sum_{n=0}^{\infty}C^{1}_{g,n+1}(t_{1})\Big(\cos\frac{\varphi^{'}}{2}|e,n+1\rangle + \sin\frac{\varphi^{'}}{2}e^{i \psi^{'}}|i,n+1\rangle\Big).
	\eeq
	To avoid the tedious calculations, $\ds\Delta_{1} = \Delta_{2} = \Delta$ and using the initial condition (\ref{cond2}), one obtains the solution of (\ref{c.gie2}) as,
	\beq
	\ba{ll}\label{second}
	\ds C^{2}_{e,n}(t) &=\frac{A_{2}g_{1}}{(g_{1}^{2}+g_{2}^{2})\sqrt{n + 1}}\left\{\mu + \left(\frac{i}{2}\Delta\sin\mu t-\mu\cos\mu t\right)e^{\frac{i}{2}\Delta t}\right\}\\
	& + \sin\frac{\varphi^{'}}{2}e^{i\psi^{'}}\eta_1~ C_{g,n}(t_{1}),\\
	\ds C^{2}_{i,n}(t)& =\frac{A_{2}g_{2}}{(g_{1}^{2}+g_{2}^{2})\sqrt{n + 1}}\left\{\mu + \left(\frac{i}{2}\Delta\sin\mu t-\mu\cos\mu t\right)e^{\frac{i}{2}\Delta t}\right\}\\
	&+ \cos\frac{\varphi^{'}}{2}\eta_1~ C_{g,n}(t_{1}),\\
	C^{2}_{g,n+1}(t)& = A_{2}ie^{-\frac{i}{2}\Delta t} \sin\mu t,
	\ea
	\eeq
	where
	\beq
	\ba{l}\label{amu2}
	\ds\mu =\frac{1}{2}\sqrt{\Delta^{2} + 4(g_{1}^{2} + g_{2}^{2})(n + 1)},\\
	\ds A_{2}(n) =-\f{1}{\mu}\left(g_{1}\sin\frac{\varphi^{'}}{2}e^{i\psi^{'}}+g_{2}\cos\frac{\varphi^{'}}{2}\right)\sqrt{n + 1}~\eta_1~ C_{g,n}^1(t_{1}).
	\ea
	\eeq
	From Eqs. (\ref{second}) and (\ref{amu2}), we obtain $C^2_{e,0}=C^2_{i,0}=C^2_{g,0}=C^2_{g,1}=0$. Therefore, $|e,0\ran$, $|i,0\ran$, $|g,0\ran$ and $|g,1\ran=0$ will not be present in Eq. (\ref{at2}) for next calculations. Using the definition of fidelity in (\ref{fidelity}), similarly, one can find the fidelity $f_2(t)$ for which the second atom collapses to the ground state by the relation
	\beq\label{fidelity2}
	f_2(t)=\f{\ds\sum\limits_{n=1}^{\infty}|C^{2}_{g,n+1}(t)|^{2}}{\ds\sum\limits_{n=1}^{\infty}\left(|C^{2}_{e,n}(t)|^{2}+|C^{2}_{i,n}(t)|^{2}+|C^{2}_{g,n+1}(t)|^{2}\right)}.
	\eeq 
	
	\subsection{Total and reduced density matrices}
	The total density matrix of the $j$ system is defined by $\rho=|\psi_j\rangle\langle \psi_j|$. Let $\rho_{A_j}$ and $\rho_{F_j}$ be the reduced density matrices of the atom and the field. Then one obtains,
	\beq
	\rho_{A_{j}}=tr_{F_j}\left(|\psi_j\rangle\langle \psi_j|\right)=\left(\ba{ccc}\rho_{A_{j}}^{ee}&\rho_{A_{j}}^{ei}&\rho_{A_{j}}^{eg}\\\rho_{A_{j}}^{ie} & \rho_{A_{j}}^{ii} & \rho_{A_{j}}^{ig}\\
	\rho_{A_{j}}^{ge} & \rho_{A_{j}}^{gi} & \rho_{A_{j}}^{gg}\ea \right),
	\eeq
	and
	\beq
	\rho_{F_j}=tr_{A_j}\left(|\psi_j\rangle\langle \psi_j|\right),
	\eeq 
	where
	\beq
	\ba{l}
	\rho_{A_{j}}^{\mu\nu}(t)=\sum\limits_{n=0}^{\infty}C_{\mu,n}^{j}(t)C_{\nu,n}^{j*}(t),~\mu,\nu=g,i,e;\\
	\rho_{A_{j}}^{\mu\nu}(t)=\left(\rho_{A_{j}}^{\nu\mu}(t)\right)^*,~
	\sum\limits_{\mu=g,i,e}\rho_{A_{j}}^{\mu\mu}(t)=1.
	\ea
	\eeq 
	Therefore, the photon number distribution $\ds p_{j}(n, t)$ is defined by \cite{pnd1}
	\beq
	\ds p_{j}(n, t) = \langle n|\rho_{F_{j}} |n\rangle,~j=1,2,~n=0,1,2,....~ .
	\eeq 
	
	\section{Preliminaries of some entropic functionals}\label{sec3.preliminaries}
	\subsection{Purity and linear entropy}  
	For a pure state, measure of entanglement is widely used whereas, for mixed state measure of entanglement, distillable entanglement and relative entropy are accepted for uncertainty measurements \cite{wootters}.
	The measure of mixedness of a quantum state with density matrix $\rho(t)$ is defined by $\mathcal{L}(t)=1-P(t)=1-Tr\left[\rho(t)^2\right]$. The quantity, $\mathcal{L}(t)=0$ for pure states and  $\mathcal{L}=\f{N-1}{N}$, for a normalized $N$-qubit quantum system with idensitity $1/N$ \cite{mixed2}. In this case, the state is called fully mixed. In particular, for a maximally entangled bipartite state \cite{entangle,entangle.2}, $\mathcal{L}(t)=0.5$. The quantity $\mathcal{L}(t)$ is known as linear entropy \cite{pipek.2.2,entangle,entangle.2,entangle.3}. It can be written as a function of the Bloch sphere radius $\eta_R(t)$ and it is defined by $\mathcal{L}=\f{1}{2}\left(1-\eta_R^2(t)\right)$. For the JCM, the linear entropy has been defined in the text book \cite{entangle.3}. Moreover, $P(t)$ describes the purity of a quantum system \cite{pati}. If $P(t)=1$, then, the system is a pure state and if $P(t)<1$, then the system is a mixed state. In this paper, we will find the purity and the linear entropy of the reduced density matrices of two quantum systems. The purity of the reduced density matrix $\rho_{A}$ is defined by
	\beq
	P(t)=\sum\limits_{\mu=g,i,e}\left(\rho_{A}^{\mu\mu}\right)^2+2\left(|\rho_{A}^{ei}|^2+|\rho_{A}^{eg}|^2+|\rho_{A}^{ig}|^2\right).
	\eeq 
	The quantity $P(t)$ is also used as a measure of disequilibrium of a quantum system and it is defined as the average position of the density function $\rho_{A}(t)$. 
	This $\mathcal{L}$ defines the measure of impurity of a quantum system. Moreover, it has been used as a measure of decoherence \cite{decoherence}, entanglement \cite{entangle,entangle.2}, complexity \cite{complexity} and mixedness \cite{mixed,mixed2}.
	\subsection{von-Neumann entropy}
	A standard measure of mixedness of a quantum state is the von-Neumann entropy and it is defined by  \cite{v1} 
	\beq
	\ds \mathcal{N} = -Tr[~\rho \ln \rho],
	\eeq
	where $\rho$ is the density operator. Then, $\mathcal{N}= 0$ for a pure state and $\mathcal{N} > 0$ for a mixed state. As the entropy is a measure of purity of states, it does not differentiate between pure states. When the state  initially is a pure state, the system remains in a pure state at any time and its entropy is always zero. The von-Neumann entropies $\mathcal{N}_{A}$ and $\mathcal{N}_{F}$ of an atom and a field are defined by
	\beq
	\ba{l}
	\ds \mathcal{N}_{A} = -Tr_{A}[~\rho_{A} \ln \rho_{A}],~
	\ds \mathcal{N}_{F} = -Tr_{F}[~\rho_{F} \ln \rho_{F} ],
	\ea
	\eeq
	where $\rho_{A}$ and $\rho_{F}$ are the reduced density matrices of an atom $A$ and a field $F$ respectively. From, the Araki and Lieb theorem \cite{lieb}, one obtains
	\beq
	\ds|\mathcal{N}_{A}- \mathcal{N}_{F} | \le \mathcal{N} \le |\mathcal{N}_{A} + \mathcal{N}_{F} |,
	\eeq
	where $\mathcal{N}$ is the total entropy of the complete atom-field system. If $\mathcal{N} = 0$, then one obtains $\ds \mathcal{N}_{A} = \mathcal{N}_{F}$. If $\lambda_{A,i}$, \(i=1,2,3\) are the eigenvalues of the density operator $\rho_A$, then von-Neumann entropy can be written as
	\beq
	\mathcal{N}_A=-\sum\limits_{i}\lam_{A,i}\ln\lam_{A,i}
	\eeq  
	\subsection{Shannon entropy}
	The temporal evolution of the information about dynamics of a field is measured by the Shannon entropy \cite{S1}. It depends on the diagonal elements of the reduced field density matrix. We can not measure the information of the phase. The Shannon entropy is associated with the classical probability distribution. 
	Therefore, the Shannon entropy of the photon number distribution $\ds p_{j}(n, t)$ is defined by \cite{pln.aop,orlowski}
	\beq\label{shannon.entropy}
	\ds \mathcal{H}(t) = -\sum_{n=0}^{\infty}~p_j(n, t) \ln p_j(n, t).
	\eeq
	
	\subsection{Atomic Wehrl entropy}
	A different definition, which clearly exhibits the unique character of coherent states, has been introduced by Wehrl \cite{w1}. In this section, we will define the atomic Wehrl entropy. To do, let us first define the atomic coherent state of a three-level atom \cite{acs,atomic.wehrl,zyczkowski.physicaE}, 
	\beq
	\ds|\theta,\Phi\rangle=\sum\limits_{\mu=e,i,g}k_{\mu}|\mu\rangle,
	\eeq
	where 
	\beq
	\ds k_{e}=\cos^{2}\frac{\theta}{2},~\ds k_{i}=\sqrt{2}\cos\frac{\theta}{2}\sin\frac{\theta}{2}e^{i\Phi},~\ds k_{g}=\sin^{2}\frac{\theta}{2}e^{2i\Phi}.
	\eeq
	It has the following properties
	\beq
	\ba{l}
	\langle \theta,\Phi|\theta,\Phi\rangle=1,~
	\ds\f{3}{4\pi}\ds\int_{\phi=0}^{2\pi}d\phi\int_{\theta=0}^{\pi}\sin\theta d\theta |\theta,\Phi\rangle\langle \theta,\Phi|=I_3,
	\ea 
	\eeq 
	where $I_3$ is the identity matrix of order 3. Then the Husimi $Q$-function of an atom is defined by
	\beq
	\ds Q(t,\theta,\Phi) =\langle \theta,\Phi|\rho_{A}|\theta,\Phi\rangle=\sum_{\mu\nu}~k_{\mu}^{*}k_{\nu}~\rho_{A}^{\mu\nu},~\ds k_{\mu\nu}= k_{\mu}^{*}k_{\nu}.
	\eeq
	The function $Q$ is also called the atomic quasiprobability distribution function and 
	the normalization condition is given by
	\beq
	\ds\f{3}{4\pi}\ds\int_{0}^{2\pi}\int_{0}^{\pi}Q(\tau,\theta,\Phi) \sin\theta ~d\theta~ d\Phi = 1.
	\eeq
	In this case, the atomic Wehrl entropy of the Husimi $Q$ function is defined by \cite{zyczkowski.physicaE,Wehrl.def,Wehrl.def.2,Wehrl.def.3,wehrl.def2,wehrl.def2.2}
	\beq
	\mathcal{W}=-\ds\f{3}{4\pi}\ds\int_{0}^{2\pi}d\Phi\int_{0}^{\pi}\sin\theta ~d\theta \,Q(\tau,\theta,\Phi) \ln Q(\tau,\theta,\Phi).
	\eeq 
	The Wehrl entropy is minimum for coherent state and maximum for the delocalized states \cite{zyczkowski.physicaE} and it corresponds to the maximally mixed state. 
	\subsection{R\'enyi entropy}
	The one-parameter R\'enyi entropy of order $\bar{\a}>0,~\ne 1$ of $\mathcal{P}$ is defined by \cite{Renyi}
	\beq\label{Renyi}
	\ba{ll}
	\mathcal{R}_{\mathcal{P}}^{(\bar{\a})}&=\f{1}{1-\bar{\a}}\ln\left[\ds\sum\limits_{i=1}^np_i^{\bar{\a}}\right],\mbox{if~}\mathcal{P}=(p_1,p_2,...,p_n),\mbox{is~discrete},\\\\
	&=\f{1}{1-\bar{\a}}\ln\left[\ds\int_{\Om^D} \left[p(x)\right]^{\bar{\a}} dx\right],\mbox{if~}\mathcal{P}=(p(x):x\in M^D),\\
	&~\mbox{is~continuous},
	\ea 
	\eeq
	where $\Om^D\subset M^D$ is the effective domain of $\mathcal{P}$ \cite{IJQC.local,dn.rcr}. The R\'enyi entropy (\ref{Renyi}) can be redefined by
	\beq\label{Renyi.dis}
	\ba{ll}
	\mathcal{R}_{\mathcal{P}}^{(\bar{\a})}&=\f{1}{1-\bar{\a}}\ln\sum\limits_{i} p_i^{\bar{\a}},~\mathcal{P}=(p_1,p_2,...,p_n),\\\\
	&=\f{1}{1-\bar{\a}}\ln\sum\limits_{i} p_i^{\bar{\a}}\mathbb{L}(\Om_i),~\mathcal{P}=(p(x):x\in M^D),
	\ea
	\eeq
	where
	\beq
	\ba{l}
	\ds\sum\limits_{i} p_i\mathbb{L}(\Om_i)=1,~0\le p_i\le 1,\mathbb{L}(\Om_i)\ge 0,\\
	p(x)=\ds\sum\limits_{i}p_i\,\chi_{\Om_i}(x),\\
	\ba{ll}
	\chi_{\Om_i}(x)&=1,x\in\Om_i,\\
	&=0,x\notin\Om_i,
	\ea
	\ea 
	\eeq 
	$\Om^D=\cup_{i}\Om_i$ and $\mathbb{L}(\Om_i)$ is the Lebesgue measure of the set $\Om_i$. The R\'enyi entropy of a continuous random variable can expressed as a limit of a sum (37) and the corresponding integation can be considred as an example of Lebesgue integration. Then the sum (\ref{Renyi.dis}) is a good approximation of (\ref{Renyi}), if the partition point set $\Om_i$ has infinitesimal Lebesgue measure, that is $\mathbb{L}(\Om_i)\rightarrow 0$, for continuous distribution. The R\'enyi entropy of a finite dimensional discrete distribution is always positive and it may be negative for continuous distribution. Another generalized entropy, of order $\bar{\a}>0,~\ne 1$ is defined by
	\beq
	\mathcal{T}_{\mathcal{P}}^{(\bar{\a})}=\f{1}{1-\bar{\a}}\left(e^{(1-\bar{\a})\mathcal{R}_{\mathcal{P}}^{(\bar{\a})}}-1\right),~ \bar{\a}>0,~ \ne 1.
	\eeq
	The entropy $\mathcal{T}_{\mathcal{P}}^{(\bar{\a})}$ is called the Tsallis entropy \cite{tsallis}. If $\mathcal{P}$ is a delta function, then $\mathcal{R}_{\mathcal{P}}^{(\bar{\a})}=\mathcal{T}_{\mathcal{P}}^{(\bar{\a})}=0$. In the limiting case, $\bar{\a}\rightarrow1$, the R\'enyi and Tsallis entopies reduce to the Shannon entropy. Moreover, if $\mathcal{P}$ is a Husimi Q function and $\bar{\a}\rightarrow 1$, then, the R\'enyi entropy reduces to an atomic Wehrl entropy. For two special cases, $\bar{\a}=0$ and $\bar{\a}=\infty$, we obtain \cite{IJQC.local,dn.rcr,Renyi.stat.2}
	\beq
	\ba{ll}
	\mathcal{R}_{\mathcal{P}}^{(0)}&=\ln n,~\mathcal{P}=(p_1,p_2,...,p_n),\\
	\mathcal{R}_{\mathcal{P}}^{(0)}&=\ln\mathbb{L}(\Om^D), \mathcal{P}=(p(x):x\in M^D),
	\ea
	\eeq
	and  
	\beq
	\ba{ll}
	\lim\limits_{\bar{\a}\rightarrow\infty}\mathcal{R}_{\mathcal{P}}^{(\bar{\a})}&=-\ln \max\limits_{i}p_i,~\mathcal{P}=(p_1,p_2,...,p_n),\\
	&=-\ln \sup\limits_{{x\in\Om^D}}p(x),~\mathcal{P}=(p(x):x\in M^D).
	\ea 
	\eeq   
	\subsection{Majorization}
	Let $\mathcal{P}=(p(1),p(2),\dots,p(n))$ be a finite discrete probability distribution (FDD). Then the cumulative distribution of $\mathcal{P}$ is defined by 
	\beqas
	\ba{ll}
	cp(k)=\sum\limits_{j=1}^{k} p(j),~ k=1,2,\dots,<n,~
	cp(n)=1.
	\ea
	\eeqas 
	If $\mathcal{P}=(p(1),p(2),\dots,p(n),\dots )$ is an infinite discrete probability distribution, then we can write
	\beqas
	\ba{l}
	cp(k)=\sum\limits_{j=1}^{k} p(j),~ k=1,2,\dots,<\infty,~
	cp(\infty)=1.
	\ea
	\eeqas 
	Let $\mathcal{P}\left(p:I\rightarrow[0,1]\right)$ be a continuous probability density function defined on $I\subset\mathbb{R}^r$. Then the cumulative distribution of $\mathcal{P}$ is defined by
	\beqas
	\ba{ll}
	cp(I_x)=\ds\int_{I_x}p(y)\,d\mu,~I_x\subset I,~
	cp(I)=1.
	\ea
	\eeqas 
	The measure $d\mu$ depends on the domain of definition of $\mathcal{P}$. 
	Let $\mathcal{P}_1=(p_1(1),p_1(2),\dots,p_1(n))$ and $\mathcal{P}_2=(p_2(1),p_2(2),\dots,p_2(n))$ be two finite discrete distributions. Then $\mathcal{P}_1$ majorizes $\mathcal{P}_2$ $(\mathcal{P}_1\succ\mathcal{P}_2)$, if they satisfy the relations \cite{majorization}
	\beqas
	\ba{l}
	cp_1^{\downarrow}([k])\ge  cp_2^{\downarrow}([k]),~1\le k<n,~
	cp_1^{\downarrow}([n])=cp_2^{\downarrow}([n])=1,
	\ea 
	\eeqas
	where $p_j[1]=\max\left\{p_j(i):i=1,2,\dots,n\right\},~p_j[n]=\min\left\{p_j(i):i=1,2,\dots,n\right\},~p_j[1]\ge p_j[2]\ge ...\ge p_j[n]$ for $j=1,2$.
	
	If $\mathcal{S}$ is a convex entropic functional of density functions, then $\mathcal{P}\succ \mathcal{Q}\implies\mathcal{S}(\mathcal{P})\ge \mathcal{S}(\mathcal{Q})$ holds \cite{w1,majorization.open,Lewenstein}.
	\noindent If $\mathcal{S}$ is a concave entropic functional of density functions, then $\mathcal{P}\succ \mathcal{Q}\implies\mathcal{S}(\mathcal{P})\le \mathcal{S}(\mathcal{Q})$ holds \cite{w1,majorization.open,Lewenstein}.\\ 
	As we know that the Shannon entropy $\mathcal{H}$ is a concave entropic functional, therefore, $\mathcal{P}\succ \mathcal{Q}\implies\mathcal{H}(\mathcal{P})\le \mathcal{H}(\mathcal{Q})$.\\ 
	The entropic moment $\mathcal{E}_{\bar{\a}}(\mathcal{P})=\ds\sum\limits_{i=1}^np_i^{\bar{\a}}$ is a concave or convex funcational, if $0<\bar{\a}<1$ or $\bar{\a}>1$, therefore, 
	\beqas
	\mathcal{P}\succ \mathcal{Q}\implies\left\{\ba{ll}\mathcal{E}_{\bar{\a}}(\mathcal{P})\le \mathcal{E}_{\bar{\a}}(\mathcal{Q})&\mbox{if~}0<\a<1\\\mathcal{E}_{\bar{\a}}(\mathcal{P})\ge \mathcal{E}_{\bar{\a}}(\mathcal{Q})&\mbox{if~}\bar{\a}>1\ea\right. .
	\eeqas 
	Consquently, the R\'enyi entropy, $\mathcal{R}_{\mathcal{P}}^{(\bar{\a})}=\ds\f{1}{1-\bar{\a}}\ln\left[\mathcal{E}_{\bar{\a}}(\mathcal{P})\right]$ is a concave functional for any $\bar{\a}>0$. Hence, $\mathcal{P}\succ \mathcal{Q}\implies\mathcal{R}_{\mathcal{P}}^{(\bar{\a})}\le \mathcal{R}_{\mathcal{Q}}^{(\bar{\a})}$ for all $\bar{\a}>0$.\\
	Similarly, the R\'enyi and atomic Wehrl entropies are concave entropic functional of Husimi $Q$-functions. Therefore, for two Husimi $Q$-functions $\mathcal{Q}_1$, $\mathcal{Q}_2$, such that $\mathcal{Q}_1\succ \mathcal{Q}_2$, then $\mathcal{R}_{\mathcal{Q}_1}^{(\bar{\a})}\le \mathcal{R}_{\mathcal{Q}_2}^{(\bar{\a})}$, for all $\bar{\a}>0$ and $\mathcal{W}({\mathcal{Q}_1})\le \mathcal{W}({\mathcal{Q}_2})$.
	\noindent Contrapositively, if $\mathbb{S}$ is a set of convex functionals of discrete distributions and $\mathcal{P}(p)$ and $\mathcal{Q}(q)$ are two discrete distributions which satisfy the relation $\sum\limits_{i}\mathcal{S}(p(i))\ge \sum\limits_{i}\mathcal{S}(q(i))$, for all $\mathcal{S}\in \mathbb{S}$, then $\mathcal{P}\succ\mathcal{Q}$.\\
	\noindent Similarly, if $\mathbb{S}$ is a set of concave functionals of discrete distributions and $\mathcal{P}(p)$ and $\mathcal{Q}(q)$ are two discrete distributions which satisfy the relation  
	$\sum\limits_{i}\mathcal{S}(p(i))\le \sum\limits_{i}\mathcal{S}(q(i))$ for all $\mathcal{S}\in \mathbb{S}$, then $\mathcal{P}\succ \mathcal{Q}$.\\
	The localization effect on R\'enyi entropy and the generalized R\'enyi complexity has been shown in our previous studies \cite{IJQC.local,dn.rcr}. 
	
	Let $\mathcal{P}_1\left(p_1:I\rightarrow[0,1]\right)$ and $\mathcal{P}_2\left(p_2:I\rightarrow[0,1]\right)$ be two continuous probability density functions defined on $I\subset\mathbb{R}^r$. Then $\mathcal{P}_1$ majorizes $\mathcal{P}_2$ $\left(\mathcal{P}_1\succ \mathcal{P}_2\right)$, if \cite{joe} 
	\beqas
	\ba{l}
	\ds\int\left[p_1(x)-r\right]^+d\mu\ge\ds\int\left[p_2(x)-r\right]^+d\mu,~\mbox{holds,~for~all~} r\ge 0,
	\ea
	\eeqas 
	where $\left[y\right]^+=\max\left\{y,0\right\}$.

	Let $f$ and $g$ be two continuous density functions, then $\ds\int \left[f-t\right]^+d\mu\le \int \left[f-t\right]^+d\mu$ holds for all $t\ge 0$ if and only if $\ds\int \phi(f)\,d\mu\le \int \phi(g)\,d\mu$, holds for all real valued convex continuous function $\phi$, such that $\phi(0)=0$ and the integral exist, where $\left[y\right]^+=\max\{y,0\}$ \cite{joe}.
	
	Let $f$ and $g$ be two continuous density functions, then $\ds\int \left[f-t\right]^+d\mu\le \int \left[f-t\right]^+d\mu$ holds for all $t\ge 0$ if and only if $\ds\int \phi(f)\,d\mu\ge \int \phi(g)\,d\mu$, holds for all real valued concave continuous function $\phi$, such that $\phi(0)=0$ and the integral exist, where $\left[y\right]^+=\max\{y,0\}$ \cite{joe}.
	
	A region $\Om\subset I\subset \mathbb{R}^r$ is called the effective domain of the density function $\mathcal{P}(p:I\rightarrow [0,1])$, if \cite{IJQC.local,dn.rcr} $\ds\int_{\Om} p(x)d\mu=1$ and for $\Om_1\subset I$, such that $\ds\int_{\Om_1}p(x)d\mu=1$, then $\mathbb{L}(\Om)\le\mathbb{L}(\Om_1)$.
	Let $\mathcal{P}_1\left(p_1:I\rightarrow[0,1]\right)$ and $\mathcal{P}_2\left(p_2:I\rightarrow[0,1]\right)$ be two continuous probability density functions and $\Om_1$, and $\Om_2$ be their respective effective domains. Then, $\mathcal{P}_1$ is called localized from $\mathcal{P}_2$, if \cite{IJQC.local,dn.rcr} $\mathbb{L}(\Om_1)\le\mathbb{L}(\Om_2)$.
	It is to be noted that, if the number of zeros of $p_2$ is greater than the number of zeros of $p_1$, then $\mathbb{L}(\Om_1)\le\mathbb{L}(\Om_2)$. Now, we have investigated a relation between majorization and localization properties for continuous distributions.
	
	\textbf{Theorem:}\label{Th.local.majo}
	Let $\mathcal{P}_1\left(p_1:I\rightarrow[0,1]\right)$ and $\mathcal{P}_2\left(p_2:I\rightarrow[0,1]\right)$ be two density functions, such that $\mathcal{P}_1\succ\mathcal{P}_2$, then $\mathcal{P}_1$ is localized from $\mathcal{P}_2$.
	
	{\bf Proof}. Since, $\mathcal{P}_1\succ\mathcal{P}_2$, therefore, 
	\beq
	\ds\int\left[p_1(x)-r\right]^+d\mu\ge \ds\int\left[p_2(x)-r\right]^+d\mu,~ \mbox{for~ all}~ r\ge 0.
	\eeq
	Then for each $r\ge 0$, there exists $I_1(r),~I_2(r)\subset I$, such that,\\ $\ds\int\left[p_1(x)-r\right]^+d\mu=\ds\int_{I_1}(p_1(x)-r)d\mu=\ds\int_{I_1}\,p_1(x)\,d\mu-r\mathbb{L}(I_1)\le 1$, and\\ $\ds\int\left[p_2(x)-r\right]^+d\mu=\ds\int_{I_2}(p_2(x)-r)d\mu=\ds\int_{I_2}\,p_2(x)\,d\mu-r\mathbb{L}(I_2)\le 1$.\\
	In the limiting case $(r\rightarrow 0+)$, we obtain 
	\beq
	\ba{l}\label{theorem}
	\ds\lim\limits_{r\rightarrow 0+}\int\left[p_1(x)-r\right]^+d\mu = 1,~
	\ds\lim\limits_{r\rightarrow 0+}I_1(r) = \Om_1,\\ 
	\ds\lim\limits_{r\rightarrow 0+}\int\left[p_2(x)-r\right]^+d\mu = 1,~
	\ds\lim\limits_{r\rightarrow 0+}I_2(r) = \Om_2,
	\ea
	\eeq
	where $\Om_1$ and $\Om_2$ are respectively effective domain of $p_1$ and $p_2$.\\ 
	If $r>1$, then $\ds\int\left[p_1(x)-r\right]^+d\mu=\ds\int\left[p_2(x)-r\right]^+d\mu=0$.\\
	Therefore, $\ds\int_{I_1(r)}\,p_1(x)\,d\mu-r\mathbb{L}(I_1(r))\ge \ds\int_{I_2(r)}\,p_2(x)\,d\mu-r\mathbb{L}(I_2(r))$, for $0\le r\le 1$,\\
	$\ds\int_{I_1(r)}\,p_1(x)\,d\mu-\ds\int_{I_2(r)}\,p_2(x)\,d\mu+r\left[\mathbb{L}(I_2(r))-\mathbb{L}(I_1(r))\right]\ge 0$, for $0\le r\le 1$.\\
	Taking $\lim\limits_{r\rightarrow 0+}$ on both sides and using (\ref{theorem}) we obtain $\lim\limits_{r\rightarrow 0+}r\left[\mathbb{L}(I_2(r))-\mathbb{L}(I_1(r))\right]\ge 0$.\\
	If possible let $\mathbb{L}(\Om_2)-\mathbb{L}(\Om_1)<0$, then there exists $\delta>0$ such that $\mathbb{L}(I_2(r))-\mathbb{L}(I_1(r))<0$ for all $r\in(0,\delta)$. Therefore, $r\left[\mathbb{L}(I_2(r))-\mathbb{L}(I_1(r))\right]<0$ for all $r\in(0,\delta)$. This implies that $\lim\limits_{r\rightarrow 0+}r\left[\mathbb{L}(I_2(r))-\mathbb{L}(I_1(r))\right]< 0$ and it will contradict that $\mathcal{P}_1\succ\mathcal{P}_2$.\\
	Hence $\mathbb{L}(\Om_1)\le \mathbb{L}(\Om_2)$, that is, $\mathcal{P}_1$ is localized from $\mathcal{P}_2$.\\
	{\bf Note:} We can not conclude about the converse of this theorem. For example, in one dimensional particle in a box of length $d$, we can say that each density function $p_n(x)=\sqrt{\frac{d}{2}}\,\sin\left(\frac{(n+1)\pi x}{d}\right)$, $0\le x\le d$, $(n=0,1,2,,\cdots)$ has effective domain $[0,~ d]$ whose Lebesgue measure is $d$. In particular, for noncentral Kratzer potential we have been prove that the narrowest confined and widest spread radial wave functions dominate the localization property of rotational wave functions for the optimum measure of R\'enyi entropy. In addition, we have been found that, the minimum and the maximum values of the R\'enyi entropy for the narrowest confined and widest spread radial wave functions, respectively \cite{IJQC.local}. 
	\section{Applications of entropic functionals to an atom-field interaction system}\label{sec4.application}
	\subsection{Fidelity, purity and linear entropy}
\begin{figure}[h]
		\centering	
		\includegraphics[width=9cm,height=7cm]{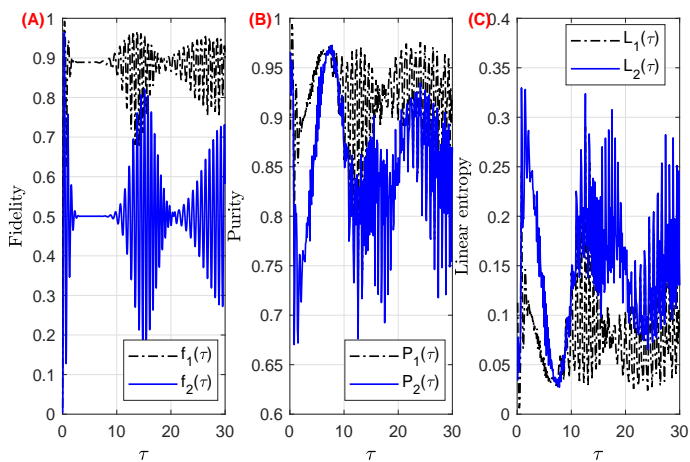}
		\caption{\label{Fig2.Fidelity} Plot of (A) the values of the fidelity's for which the first (dot dashed) and second (solid) atom collapse to the ground state. (B) The purity of the reduced density matrices of the first (dot dashed) and second (solid) atoms. (C) Linear entropy of the first (dot dashed) and the second (solid) atoms. The parameters are $g_{1}=g_{2}=g$, $\D=0$, $\psi=0$, $\varphi=\frac{\pi}{2}$, $\psi^{'}=0$, $\varphi^{'}=\frac{\pi}{2}$, $\tau_{1}=0.137503$, $\zeta=0$ and $\bar{n}=8$.}
	\end{figure}

	For numerical calculations, the driving field is a coherent electromagnetic field \cite{pnd1} have been considered. Therefore, coefficients of number state $|n\rangle$ are defined by $\ds C_{n} = e^{-\frac{|\alpha|^{2}}{2}}\frac{\alpha^{n}}{\sqrt{n!}}$, $\ds\a=\sqrt{\bar{n}}~e^{i\zeta}$, \(n=0,1,2,\cdots\). The quantity $\bar{n}$, is called mean photon number of coherent field and $\zeta$ is phase of $\alpha$. In addition, 
	\beq\label{conditions}
	\ba{l}
	\D=\psi=\psi^{'}=0,~ \varphi=\varphi'=\frac{\pi}{2},g_{1}=g_{2}=g,
	\ea 
	\eeq
	have been considered. Then, first atom collapses to the ground state at $\ds\tau_1= 0.137503$, with fidelity $f_1(\tau_1)=0.990682$, which is the maximum value of $f_1$, for a coherent field having mean photon number $\bar{n} = 8$. With the help of Eqs. (\ref{second}) and (\ref{amu2}), we observe that, the second atom-field interacting system is same as that of first atom-field interacting system, if $f_1$ is equal to one. But in the real system the fidelity is always less than one. Therefore, the maximum fidelity is considered in this paper. 
	The fidelity's for which the first and the second atoms collapse to the ground state are plotted in Fig. \ref{Fig2.Fidelity} (A). From this, it is observed that, the first atom is stable to the ground state but the second atom is a superposition of $|e\rangle$, $|i\rangle$ and $g\rangle$.
	The purity and the linear entropy of states of first-atom and second-atom field interacting systems are displayed in Figs. \ref{Fig2.Fidelity} (B) and \ref{Fig2.Fidelity} (C), under the conditions (\ref{conditions}). The second atom field interacting system is always a mixed state. From, this figure, it is observed that, purity of first system is greater than that of the second system. 
	\subsection{von-Neumann entropy}
	The density matrix $\rho_{A_{1}}$ depends on $t$ and the density matrix $\rho_{A_{2}}$ depends on time $t_{1}$ and $t$. The von-Neumann entropy of the $j$th atom is defined by
	\beq
	\ds \mathcal{N}_j = -\sum_{i=1}^{3}\lambda_{i}^{j}\ln\lambda_{i}^{j},~j=1,2,
	\eeq
	where $\lambda_{i}^{j}~(i=1,2,3)$ are eigenvalues of $\rho_{A_{j}}$. Therefore, $\lambda_{i}^j(i=1,2,3)$ are the roots of the equation
	\beq
	\ba{l}
	x^3-x^2+x\left[\rho_{A_j}^{ee}\rho_{A_j}^{ii}+\rho_{A_j}^{ee}\rho_{A_j}^{gg}+\rho_{A_j}^{ii}\rho_{A_j}^{gg}-|\rho_{A_j}^{ei}|^2\right.\\
	\left.-|\rho_{A_j}^{eg}|^2-|\rho_{A_j}^{ig}|^2\right]-det(\rho_{A_j})=0~\mbox{for~}j=1,2.
	\ea
	\eeq 
	The explicit form of the von-Neumann entropy of the $j$th system is obtained by
	\beq\label{von-Neumann}
	\ds \mathcal{N}_j =-\lam_+^{j}\ln \lam_+^j-\lam_-^j\ln\lam_-^j,~j=1,2,
	\eeq
	where two sets of eigen values $\mathcal{P}_1^{\downarrow}=(\lam_+^1,\lam_-^1,0)$ and $\mathcal{P}_2^{\downarrow}=(\lam_+^2,\lam_-^2,0)$ in increasing order are defined by 
	\beq
	\lam_{\pm}^j=\f{1}{2}\left[1\pm  \sqrt{1-8\rho_{A_j}^{ee}+16(\rho_{A_j}^{ee})^2+8|\rho_{A_j}^{eg}|^2}\right].
	\eeq 
	The von-Neumann entropies of two atoms are binary entropies $\mathcal{N}_j=\mathcal{N}_j(\lam_+^j)=\mathcal{N}_j(\lam_-^j)$, $j=1,2$. The cumulative distributions can be written as
	\beq
	\ba{l}
	cp_1^{\downarrow}(1)=\lam_+^1,\,cp_1^{\downarrow}(2)=1,\,cp_1^{\downarrow}(3)=1,\\
	cp_2^{\downarrow}(1)=\lam_+^2,\,cp_2^{\downarrow}(2)=1,\,cp_2^{\downarrow}(3)=1.
	\ea 
	\eeq 
	\begin{figure}[h]
		\includegraphics[width=10cm,height=7cm]{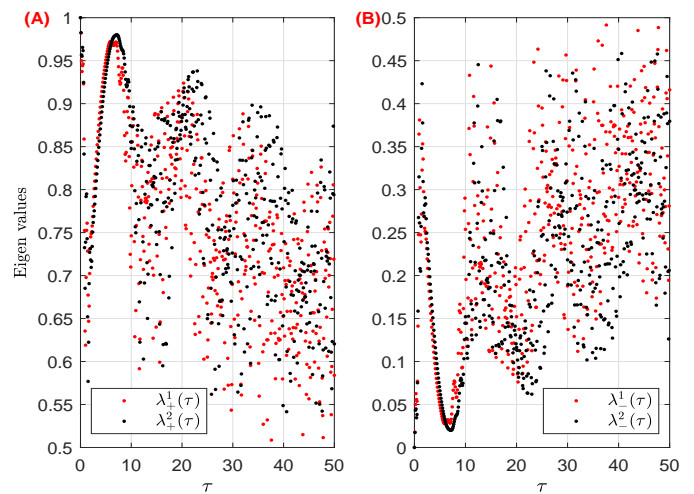}
		\caption{\label{Fig3.von_dis} (Color online) Plot of the eigen values (A) $\lam_+^1$, $\lam_+^2$ and (B) $\lam_-^1$, $\lam_-^2$. The parameters are as taken in Fig. \ref{Fig2.Fidelity}.}
	\end{figure}
	Figs. \ref{Fig3.von_dis}, display eigen values $\lam_+^{1,2}$ and $\lam_-^{1,2}$. From this figure, it is observed, that $\lam_+^1-\lam_+^2$ is indefinite in nature with respect to $\tau$. 
	\begin{figure}[h]
		\includegraphics[width=9cm,height=7cm]{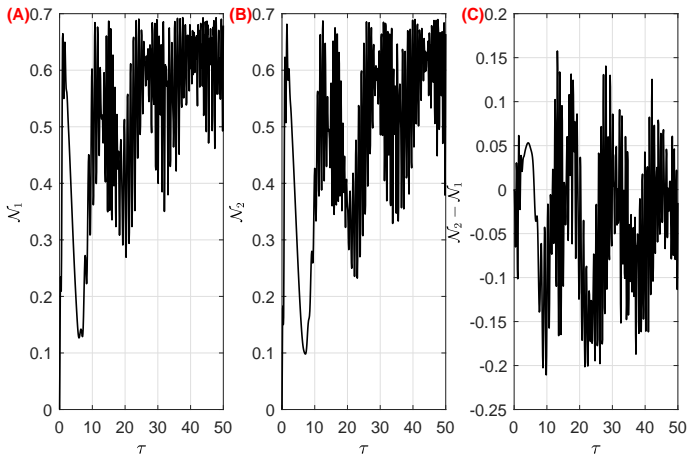}
		\caption{\label{Fig4.von} Plot of the von-Neumann entropy of the (A) 1st atom, (B) 2nd atom and (C) the difference between the von-Neumann entropies of the 2nd and the 1st atoms. The parameters are as taken in Fig. \ref{Fig2.Fidelity}.}
	\end{figure} 
	The von-Neumann entropies $\mathcal{N}_1(\tau)$ and $\mathcal{N}_2(\tau)$ are plotted Figs. \ref{Fig4.von} (A) and \ref{Fig4.von} (B) respectively, with respect to $\tau$ and difference $\ds \mathcal{N}_2(\tau)-\mathcal{N}_1(\tau)$ is plotted in Fig. \ref{Fig4.von} (C). From Fig. \ref{Fig4.von} (C), it can be conclude that, $\mathcal{N}_2>\mathcal{N}_1$ for some values of $\tau$ and $\mathcal{N}_2<\mathcal{N}_1$ for other values $\tau$. Therefore, von-Neumann entropy can not be controlled by passing atom(s) into the cavity under the conditions (\ref{conditions}).
	
	\subsection{Shannon entropy}
	The photon number distribution $p_j(n,t)$ of the field when the $j$th atom passes through the cavity is given by,
	\beq
	\ba{lll}\label{photon.dis}
	p_{1}(n,\tau)&=e^{-|a|^2}\cos^2(\sqrt{2}\,\tau),n=0,\\
	&=\f{|\a|^{2n}e^{-|\a|^2}}{n!}\Big(\cos^2(\sqrt{2n+2}\,\tau)+\f{n}{|\a|^{2}}\sin^2(\sqrt{2n}\,\tau)\Big),\\
	&~~~~~~~~~~~~~~~~~~~~~~~~~~~~~~~~~~~~~~~~~~~~~~~~~~~~ n>0,\\
	p_{2}(n,\tau)&=0,n=0\\
	&=\eta_1^2 e^{-|\a|^2}\sin^2(\sqrt{2}\tau_1)\cos^2(\sqrt{4}\tau),n=1,\\
	&=\eta_1^2e^{-|\a|^2}\left(\f{|\a|^{2n}}{n!}\sin^2(\sqrt{2n+2}\,\tau_1)\cos^2(\tau\sqrt{2n+4})\right.\\
	&\left.+\f{|\a|^{2n-2}}{(n-1)!}\sin^2(\sqrt{2n}\,\tau_1)\sin^2(\tau\sqrt{2n+2})\right),~ n>1.\\
	\ea
	\eeq
	\begin{figure}[h]
		\includegraphics[width=9cm,height=8cm]{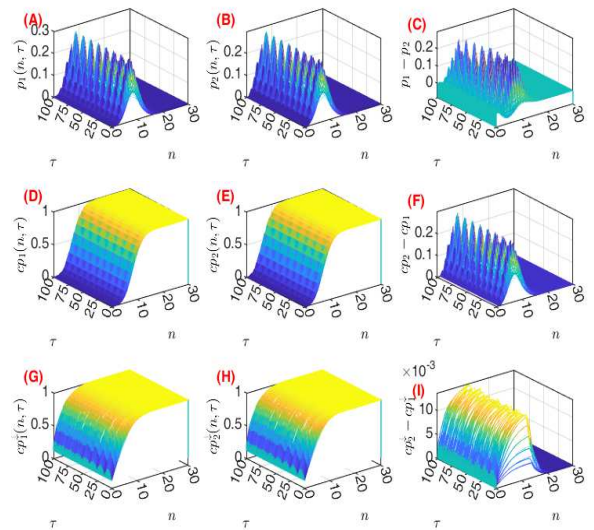}
		\caption{\label{Fig5.Majorization.Pn} (Color online) Plot of the photon number distributions of (A) 1st atom-field interaction, (B) 2nd atom-field interaction and (C) difference between $p_1$ and $p_2$. Cumulative photon number distributions of (D) 1st atom-field interaction, (E) 2nd atom-field interaction and (F) difference between $cp_2$ and $cp_1$. Cumulative photon number distributions in decreasing order of (G) 1st atom-field interaction, (H) 2nd atom-field interaction and (I) difference between $cp_2^{\downarrow}$ and $cp_1^{\downarrow}$. The parameters are as taken in Fig. \ref{Fig2.Fidelity}.}
	\end{figure}
	\noindent Figs. \ref{Fig5.Majorization.Pn} (A) and Fig. \ref{Fig5.Majorization.Pn} (B), show the photon number distributions of two fields and their difference is plotted in Fig. \ref{Fig5.Majorization.Pn} (C). From this figure it is clear that, $p_1(n,\tau)-p_2(n,\tau)$ is neither strictly positive nor strictly negative. The cumulative distributions $cp_1(n,\tau)$, $cp_2(n,\tau)$ of photon numbers are shown in Fig. \ref{Fig5.Majorization.Pn} (D), (E) respectively and the difference between them is shown in Fig. \ref{Fig5.Majorization.Pn} (F), which is strictly positive. The cumulative distributions of decreasing orders $cp_1^{\downarrow}(n,\tau)$ and $cp_2^{\downarrow}(n,\tau)$ are plotted in Figs. \ref{Fig5.Majorization.Pn} (G), (H) and their difference in Fig. \ref{Fig5.Majorization.Pn} (I). From Fig. \ref{Fig5.Majorization.Pn} (I), it can be observed that, photon number distributions of two systems satisfy the majorization condition $p_2^{\downarrow}\succ p_1^{\downarrow}$.
	
	Now, using $0\ln0=0$, and photon number distributions defined in Eq.(\ref{photon.dis}), one obtains the Shannon entropy $\mathcal{H}_j(t)$ of the field $F_j$. The Shannon entropy $(\mathcal{H}_{1}(t))$ of first atom-field interaction is a function of $t$ and $(\mathcal{H}_{2}(t))$ of second system is a function of $t_1$ and $t$. 
	\begin{figure}[h!]
		\includegraphics[width=8cm,height=7cm]{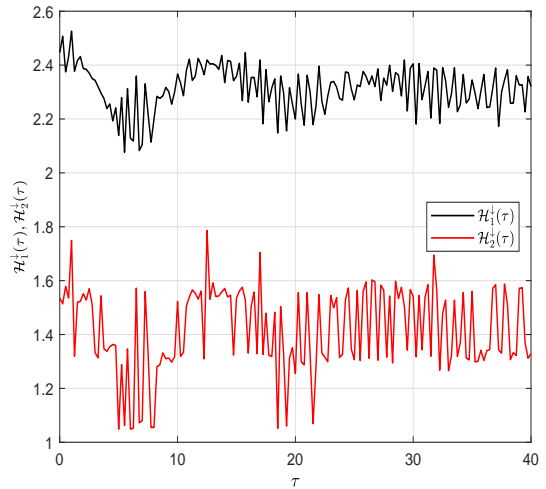}
		\caption{\label{Fig6.Shannon.Pn} Plot of the Shannon entropy of 1st atom-field interaction (black line) and 2nd atom-field interaction (red line). The parameters are as taken in Fig. \ref{Fig2.Fidelity}.}
	\end{figure}
	Then, Shannon entropies $\mathcal{H}_{1}(t)$ and $\mathcal{H}_{2}(t)$ are plotted in Fig. \ref{Fig6.Shannon.Pn} as functions of $\tau$. 
	
	From Fig. \ref{Fig6.Shannon.Pn}, one can observe that, the Shannon entropy $\mathcal{H}_{2}(t)$ of the field of second system is always less than that of $\mathcal{H}_{1}(t)$ of the field of first system. Therefore, Shannon entropy can be reduced by passing another atom(s) into the cavity in the same process. 
	
	\subsection{Atomic Wehrl entropy}
	The Husimi $Q$ function of the $j$th atom is obtained as 
	\beq\label{Husimi.Qj}
	Q_j(\tau,\theta,\phi)=\sum\limits_{i=1}^3Q_j^{(i)}(\tau,\theta,\phi),
	\eeq
	where
	\beq
	\ba{ll}
	Q_1^{(1)}(\tau,\theta,\phi)&=\f{e^{-|\a|^2}}{16}\Big(5+4\cos\theta-\cos2\theta\\
	&+2\sqrt{2}\cos\phi(2\sin\theta+\sin2\theta)\Big)\\
	&\times\sum\limits_{n=0}^{\infty}\f{|\a|^{2n}}{n!}\cos^2(\sqrt{2n+2}\,\tau),\\
	Q_1^{(2)}(\tau,\theta,\phi)&=\f{e^{-|\a|^2}}{4\sqrt{2}}\Large((1-\cos2\theta)\sin2\phi\\
	& +\sqrt{2}(2\sin\theta-\sin2\theta)\sin\phi\Large)\\
	&\times\sum\limits_{n=1}^{\infty}\f{|\a|^{2n-1}}{\sqrt{n!(n-1)!}}\cos(\sqrt{2n+2}\,\tau)\sin(\sqrt{2n}\,\tau),\\
	Q_1^{(3)}(\tau,\theta,\phi)&=\f{e^{-|\a|^2}}{8}\left(3-4\cos\theta+\cos2\theta\right)\\
	&\sum\limits_{n=0}^{\infty}\f{|\a|^{2n}}{n!}\sin^2(\sqrt{2n+2}\,\tau),
	\ea 
	\eeq
	and
	\beq
	\ba{ll}
	Q_2^{(1)}(\tau,\theta,\phi)&=\f{\eta_1^2e^{-|\a|^2}}{16}\Big(5+4\cos\theta-\cos2\theta\\
	&+2\sqrt{2}\cos\phi(2\sin\theta+\sin2\theta)\Big)\\
	&\times\sum\limits_{n=0}^{\infty}\f{|\a|^{2n}}{n!}\sin^2(\sqrt{2n+2}\,\tau_1)\cos^2(\tau\sqrt{2n+4}),\\
	Q_2^{(2)}(\tau,\theta,\phi)&=\f{\eta_1^2e^{-|\a|^2}}{4\sqrt{2}}\Big((1-\cos2\theta)\sin2\phi\\\ 
	&+\sqrt{2}(2\sin\theta-\sin2\theta)\sin\phi\Big),\\
	&\times\sum\limits_{n=1}^{\infty}\Big[\f{|\a|^{2n-1}}{\sqrt{n!(n-1)!}}\cos(\sqrt{2n+4}\,\tau)\\
	&\sin(\sqrt{2n+2}\,\tau)\sin(\sqrt{2n+2}\,\tau_1)\sin(\sqrt{2n}\,\tau_1)\Big]\\
	Q_2^{(3)}(\tau,\theta,\phi)&=\f{\eta_1^2e^{-|\a|^2}}{8}\left(3-4\cos\theta+\cos2\theta\right)\\
	&\sum\limits_{n=0}^{\infty}\f{|\a|^{2n}}{n!}\sin^2(\sqrt{2n+2}\,\tau_1)\sin^2(\tau\sqrt{2n+4}).
	\ea 
	\eeq
	Let us consider
	\beq
	\ba{ll}
	\left[Q_j(\tau,r)\right]^+=\ds\int_{\theta=0}^{\pi}\int_{\phi=0}^{2\pi}\left[Q_j(\tau,\theta,\phi)-r\right]^+\sin\theta\,d\theta d\phi,r\ge0
	\ea
	\eeq
	for $,j=1,2$.
	\begin{figure}[h]
		\includegraphics[width=9cm,height=8cm]{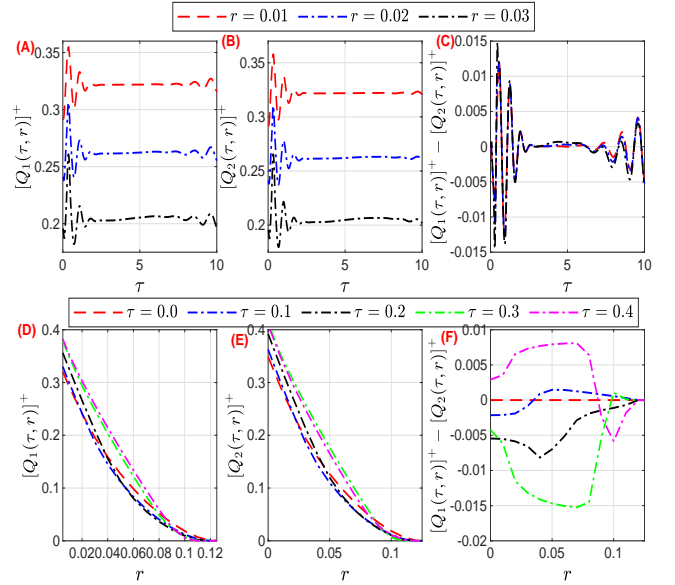}
		\caption{\label{Fig9} (Color online) In (A) $\left[Q_1(\tau,r)\right]^+$, (B) $\left[Q_2(\tau,r)\right]^+$, their difference (C) $\left[Q_1(\tau,r)\right]^+-\left[Q_2(\tau,r)\right]^+$ with respect to $\tau$. Similarly, they are plotted with respect to $r$ in (D), (E) and (F). The parameters are as taken in Fig. \ref{Fig2.Fidelity}.}	
	\end{figure}
	Now, Figs. \ref{Fig9} (A), (B) and (C), show $\left[Q_1(\tau,r)\right]^+$, $\left[Q_2(\tau,r)\right]^+$ and their differences with respect to \(\tau\) for different \(r\). From Fig. \ref{Fig9} (C), it is clear that, $\left[Q_1(\tau,r)\right]^+-\left[Q_2(\tau,r)\right]^+$ is oscillating with respect to $\tau$ for different values of $r$. 
	
	Similarly, Figs. \ref{Fig9} (D), (E) and (F), display $\left[Q_1(\tau,r)\right]^+$, $\left[Q_2(\tau,r)\right]^+$ and differences between them with respect to $r$, for different normalized time $\tau$. 
	
	We obtain, $\lim\limits_{r\rightarrow 0}\left[Q_j(\tau,r)\right]^+=1$, and  $\lim\limits_{r\rightarrow 1}\left[Q_j(\tau,r)\right]^+=0$, for $j=1,2$. Moreover, it is verified that, $\left[Q_1(\tau,r)\right]^+\gtrless\left[Q_2(\tau,r)\right]^+$, for $0\le r\le 1$ which implies \(Q_1(\tau,r)\nsucc Q_1(\tau,r)\), \(Q_1(\tau,r)\nprec Q_1(\tau,r)\) and therefore, $\implies\mathcal{W}_1(\tau)\lessgtr\mathcal{W}_2(\tau)$. 
	\begin{figure}[h]
		\includegraphics[width=9cm,height=7cm]{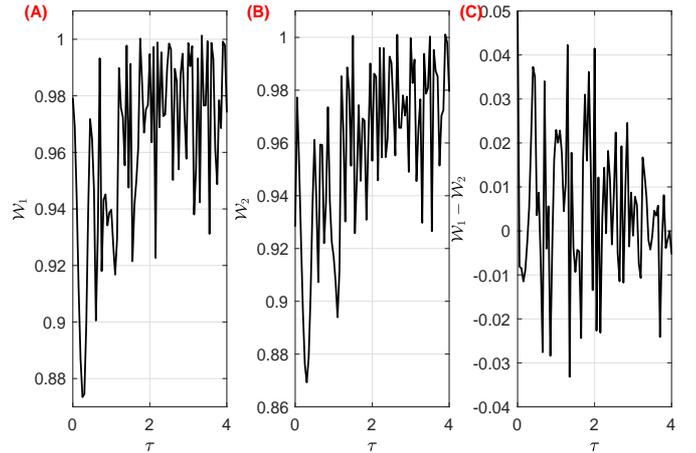}
		\caption{\label{Fig10} Plot of the atomic Wehrl entropy of (A) the 1st atom, (B) the atom and (C) the difference between them. The parameters are as taken in Fig. \ref{Fig2.Fidelity}.}	
	\end{figure}
	The atomic Wehrl entropies $\mathcal{W}_1(\tau)$ and $\mathcal{W}_2(\tau)$ are plotted with respect to $\tau$ in Fig. \ref{Fig10} (A) and \ref{Fig10} (B) respectively. Fig. \ref{Fig10} (C) shows the difference between atomic Wehrl entropies, $\mathcal{W}_2(\tau)-\mathcal{W}_1(\tau)$. 
	
	From Fig. \ref{Fig10} (C), it is observed that, difference between atomic Wehrl entropies is oscillating as $\left[Q_j(\tau,r)\right]^+-\left[Q_j(\tau,r)\right]^+$ is indefinite (\ref{Fig9} (F)) with respect to $r$ for different values of $\tau$. Therefore, atomic Wehrl entropy cannot be controlled by adding atom(s) into the cavity.
	\subsection{R\'enyi entropy}
	Using the definition (\ref{Renyi}), the R\'enyi entropy of the photon number distributions (\ref{photon.dis}) is defined by
	\beq
	\mathcal{R}_{\mathcal{P}_j}^{(\bar{\a})}(\tau)=-\f{1}{1-\bar{\a}}\ln\sum\limits_{n=n_0}^{n_{max}} \left[p_j(n,\tau)\right]^{\bar{\a}},j=1,2
	\eeq
	and for the Husimi Q function it is defined by
	\beq
	\mathcal{R}_{Q_j}^{(\bar{\a})}(\tau)=-\f{1}{1-\bar{\a}}\ln\ds\left[\int_{\theta=0}^{\pi}\int_{\phi=0}^{2\pi} \left(Q_j(\tau,\theta,\phi)\right)^{\bar{\a}}\sin\theta d\theta d\phi\right], 
	\eeq
	for $j=1,2$. 
	The R\'enyi entropy of photon number distributions $p_j(n,\tau)$ is independent of its order $p^{\downarrow}_j(n,\tau)$, i.e., $\mathcal{R}_{\mathcal{P}_j}^{\downarrow(\bar{\a})}(\tau)=\mathcal{R}_{\mathcal{P}_j}^{(\bar{\a})}(\tau)$. 
	\begin{figure}[h]
		\includegraphics[width=9cm,height=7.5cm]{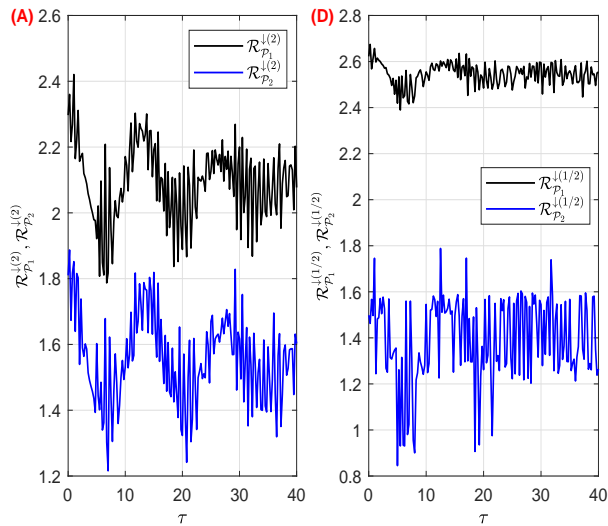}
		\caption{\label{Fig11} Plot of the R\'enyi entropy of the photon number distributions in (A) of order $\bar{\a}=2$, (B) of order $\bar{\a}=1/2$ for $p_1$ (black lines) and of $p_2$ (blue lines). The parameters are as taken in Fig. \ref{Fig2.Fidelity}.}	
	\end{figure}
	\begin{figure}[h]
		\includegraphics[width=9cm,height=8.5cm]{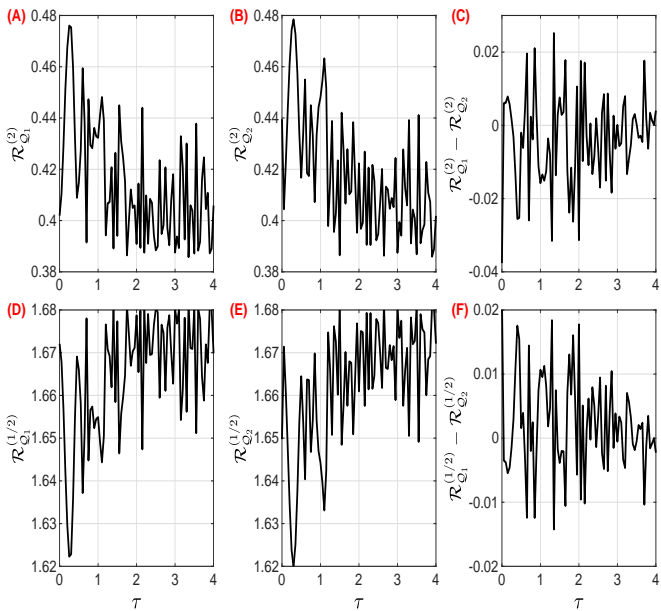}
		\caption{\label{Fig12} Plot of the R\'enyi entropy of the Husimi function in (A)-(C) of order $\bar{\a}=2$, (D)-(F) of order $\bar{\a}=1/2$ for $Q_1$ and $Q_2$. The parameters are as taken in Fig. \ref{Fig2.Fidelity}.}	
	\end{figure}
	In Fig. \ref{Fig11}, we have plotted the R\'enyi entropies $\mathcal{R}_{\mathcal{P}_1}^{\downarrow(\bar{\a})}$ and $\mathcal{R}_{\mathcal{P}_2}^{\downarrow(\bar{\a})}$ of the photon number distributions $p_1(n,\tau)$ and $p_2(n,\tau)$ for order $\bar{\a}=2$, and $1/2$. From this figure, it is clear that, $\mathcal{R}_{\mathcal{P}_2}^{\downarrow(2)}(\tau)<\mathcal{R}_{\mathcal{P}_1}^{\downarrow(2)}(\tau)$ and $\mathcal{R}_{\mathcal{P}_2}^{\downarrow(1/2)}(\tau)<\mathcal{R}_{\mathcal{P}_1}^{\downarrow(1/2)}(\tau)$. 
	Similarly, R\'enyi entropies $\mathcal{R}_{\mathcal{Q}_1}^{(\bar{\a})}$ and $\mathcal{R}_{\mathcal{Q}_2}^{(\bar{\a})}$ of the Husimi functions $Q_1$ and $Q_2$ are plotted in Fig. \ref{Fig12}. From this figure one can observe that, $\mathcal{R}_{\mathcal{Q}_1}^{(\bar{\a})}-\mathcal{R}_{\mathcal{Q}_2}^{(\bar{\a})}$ is oscillating. Since $\left[Q_1(\tau,r)\right]^+-\left[Q_2(\tau,r)\right]^+$  is oscillating hence, \(Q_1(\tau,r)\nsucc Q_1(\tau,r)\), \(Q_1(\tau,r)\nprec Q_1(\tau,r)\). 
	Therefore, the R\'enyi entropy of Husimi $Q$ function cannot be controlled by adding atom(s) into the cavity. 
	\section{Conclusions}\label{sec5.conclusion}
	In this paper, we have investigated of some entropy functionals of a vee type three-level atom which interacts with a coherent field in a resonant cavity. Two set of quantum systems (i) first atom-field interaction (ii) second atom-field interaction have been considered. Both systems have discrete (photon number distributions) and continuous probability distribution (Husimi $Q$ function). To understand the majorization effect of continuous probability density functions on entropy functionals, localization property \cite{IJQC.local,dn.rcr} is defined. It is observed that, if $\mathcal{P}\succ \mathcal{Q}$, then $\mathcal{P}$ is localized from $\mathcal{Q}$, but the converse is not true (please see theorem). Entropy functionals of second atom-field interacting system is same as that of first atom-field interacting system, if the first atom collapses to the ground state with fidelity is equal to one. The maximum fidelity and the linear entropy of two quantum systems are investigated. The first quantum system is more pure than the second quantum system. The von-Neumann, Shannon, atomic Wehrl and R\'enyi entropies are investigated for two quantum systems of photon number distributions and for quasi probability distributions. The Shannon and R\'enyi entropies of photon number distributions can be reduced by passing atom(s) into the cavity one by one. The von-Neumann entropy of atoms, atomic Wehrl and R\'enyi entropies of Husimi $Q$ functions cannot be controlled by adding atom(s) into the cavity. A similar behavior will be obtained for ladder and lambda configurations of three-level atom-field interacting system. It may be remarked that, all results are obtained for solutions of a vee-type three-level atom and single-mode quantized field. This majorization technique can be applied for solutions for Hamiltonian which is defined in Eq. (\ref{Total.Hamiltonian}). 
	

	\ed